\documentclass[runningheads]{llncs}

\usepackage[margin=0.9in]{geometry}

\usepackage{ textcomp }

\usepackage[utf8]{inputenc}
\usepackage[T1]{fontenc}

\usepackage{amssymb,amsfonts,amsmath,amsthm}
\usepackage{graphicx}
\usepackage{xcolor}

\usepackage{caption}

\usepackage{float}

\usepackage[shortlabels,inline]{enumitem}

\usepackage{xspace}
\usepackage{ifthen}

\usepackage{comment}

\usepackage{xcolor}

\PassOptionsToPackage{hyphens}{url}
\usepackage{url}

\usepackage{hyperref}
\hypersetup{
    colorlinks,
    linkcolor={blue!80!black},
    citecolor={blue!80!black},
    urlcolor={blue!80!black},
    breaklinks=true
}

\setlist[itemize]{label=\textbullet}

\begin{document}

%*-*-*-*-*-*-*-*-* [TITLE] *-*-*-*-*-*-*-*-*
\title{Does multi-block MEV exist? Analysis of 2 years of MEV Data}
%
%\titlerunning{Does multi-block MEV exist?}

%*-*-*-*-*-*-*-*-* [AUTHORS] *-*-*-*-*-*-*-*-*
\author{Pascal Stichler}

%*-*-*-*-*-*-*-* [INSTITUTE] *-*-*-*-*-*-*-*
\institute{%
ephema labs \\
\email{pascal@ephema.io}
}

%*-*-*- [SHOW TITLE + AUTHORS + INSTITUTE] *-*-*--*-*
\maketitle 

%*-*-*-*-*-*-*-* [ABSTRACT] *-*-*-*-*-*-*-*
\begin{abstract}

This study analyzes proposer-builder data and MEV-Boost payment data following the Ethereum merge in September 2022 to identify patterns of multi-block MEV. Our findings reveal fewer multi-slot sequences of builders than predicted by a random Monte Carlo simulation, with the longest observed sequence spanning 25 slots. Additionally, we observe that average MEV-Boost payments increase with the length of consecutive sequences from approximately 0.05 ETH for single slots to 0.08 ETH for nine consecutive slots. Within longer sequences, payments per slot show a slight increase, suggesting that builders bid higher for longer sequences or the first slot after a longer sequence. A weak positive autocorrelation is found between subsequent MEV-Boost payments, challenging the hypothesis of alternating periods of low and high MEV. Finally, our comparison of builders during periods of low and high base fee volatility reveals minimal correlation, indicating the absence of builder specialization based on base fee volatility. The detailed results can be found in the Jupyter notebook on \href{https://github.com/ephema/MEVBoost-Analysis/blob/762b7626c57cc6a1c350059b41e272a70cda49cf/%5Bephema%5D_MEV_Boost_Multi_Slot_MEV_Analysis.ipynb}{Github} or \href{https://colab.research.google.com/drive/1kKM-da6xP7St8puzPuyn1Ndag6a6wsg3?usp=sharing}{Google Colab}.

\end{abstract}

%*-*-*-*-*-*-*-* [BACKGROUND] *-*-*-*-*-*-*-*
\section{Background}
\label{sec:background}

Multi-block Maximal Extractable Value (MMEV) occurs when one party controls more than one consecutive block. It was first introduced in 2021 by \cite{babel2023clockwork} as k-MEV and further elaborated by \cite{mackinga2022twap}. It is commonly assumed that controlling multiple slots in a sequence allows to capture significantly more MEV than controlling them individually. This derives from MEV accruing superlinearly over time. The \href{https://collective.flashbots.net/t/multi-block-mev/457}{most discussed} multi-block MEV strategies include \href{https://eprint.iacr.org/2022/445.pdf}{TWAP oracle manipulation attacks} on DEXes and producing forced liquidations by price manipulation.

After the merge, \cite{jensen2023multi} have looked into the first four months of data on multi-block MEV and summarized it as ``\textit{preliminary and non-conclusive results, indicating [that] builders employ super-linear bidding strategies to secure consecutive block space}".

With the recent Attester-Proposer-Separation (APS) and pre-confirmation discussions, multi-block MEV has become more of a pressing issue again as it might be prohibitive for some of the proposed designs (For a more in-depth overview, we’ve created a \href{https://miro.com/app/board/uXjVK07aBCU=/?share_link_id=220296247588}{diagram of recently proposed mechanism designs} and also \href{https://x.com/mikeneuder}{Mike Neuder} lately gave a \href{https://www.youtube.com/watch?v=ToVi-zsiE4M}{comprehensive overview}).

%*-*-*-*-*-*-*-* [METHODOLOGY] *-*-*-*-*-*-*-*
\section{Methodology}
\label{sec:methodology}

In order to get a better understanding of the historical prevalence of multi-block MEV, we decided to look at all slots from the Merge in September ‘22 until May ‘24 (totalling roughly 4.3 million slots) and analyze the corresponding data on validators and builders and on MEV-boost payments (if applicable). The scope was to identify patterns of unusual consecutive slot sequences and accompanying MEV values. \href{https://mevboost.pics/data.html}{The data} has been kindly provided by Toni Wahrstätter and contains information per slot on relay, builder pubkey, proposer pubkey and MEV-Boost value as well as a builder pubkey and validator pubkey mapping. In the labeling of validators for our purposes staking pool providers such as Lido or Rocket Pool are treated as one entity.

MEV-Boost payments are used as a proxy for the MEV per block. We acknowledge that this is only a non-perfect approximation. The ascending MEV-Boost first-price auction by its nature of being public essentially functions like a second price $+1$ wei auction. Hence, we strictly speaking only get an estimate of the intrinsic value of the second highest bidder. However, as \cite{yang2024decentralization} have observed more than 88\% of MEV-Boost auctions were competitive and \cite{oz2024wins} concluded that the average profit margin per top three builder is between 1\% and 5.4\%, further indicating a competitive market between the top builders. Based on this, despite the limitations we deem it feasible to use the MEV-Boost payments as an approximation for the generated MEV per block.

To establish a baseline of expected multi-slot sequences, a Monte Carlo simulation was conducted. In this simulation, builders were randomly assigned to each slot within the specified time period, based on their observed daily market share during that period. The frequency of consecutive slots, ranging in length from 1 to 25 (the longest observed sequence in the empirical data), was recorded. This procedure was repeated 100 times, and the average was taken. We decided to use daily market shares for the main analysis as in the investigated time period market shares have strongly shifted \cite{yang2024decentralization}. For comparison we also ran the analysis on monthly and overall market shares.

Further, base fee volatility data has been included to cross-check effects of low and high-volatility periods. Previous research (e.g. \cite{gupta2023centralizing} \& \cite{heimbach2024non}) has focused on token price volatility effects based on CEX-prices. As we are interested in low- and high-MEV environments, we deem base fee volatility for our use case more fitting, as it is driven by empty or full blocks which are at least partially a result of the prevalence of MEV opportunities.

%*-*-*-*-*-*-*-* [EMPIRICAL FINDINGS] *-*-*-*-*-*-*-*
\section{Empirical Findings}
\label{sec:Empirical_Findings}

\subsection*{Finding 1: Fewer multi-slot sequences exist than assumed by random distribution}

\vspace*{-5mm}
\begin{figure}[H]
    \centering
    \includegraphics[width=0.58\linewidth]{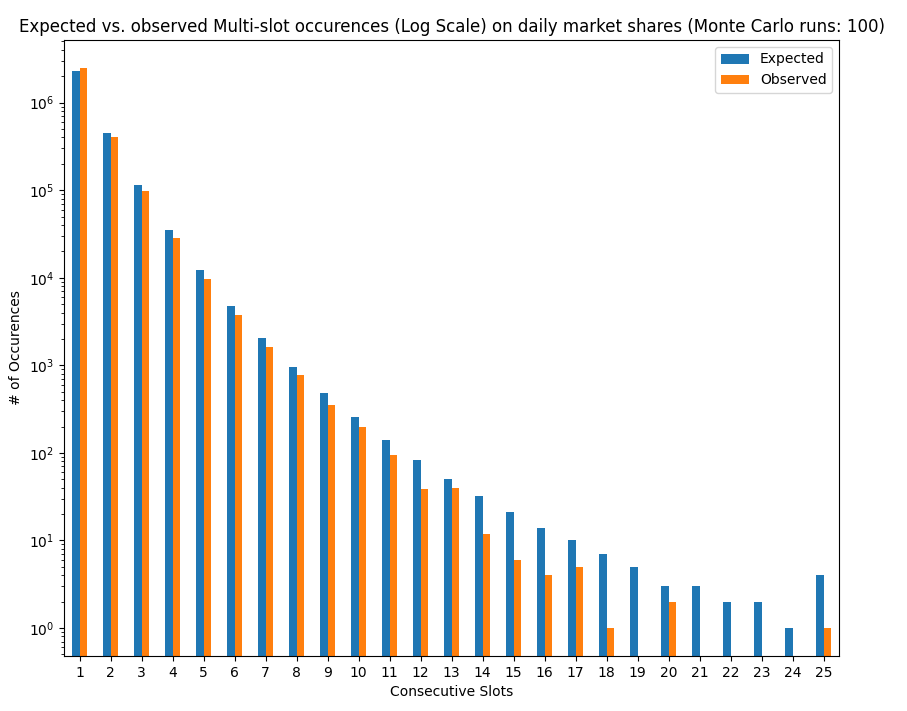}
    \caption{Comparison of statistically expected vs. observed multi-slot sequences (note that slots $>$ 25 have been summarized in slot 25 for brevity)}
    \label{Fig_1}
\end{figure}

Firstly, the prevalence of multi-slot sequences with the same builder proposing the block was investigated to determine if they are more common than would be expected by chance.

Comparing the results of the Monte Carlo simulation as a baseline in expected distribution (blue) with the observed distribution (orange), it can be seen that significantly fewer multi-slot sequences occur than expected (\autoref{Fig_1}). The longest observed sequence was 25 slots and the longest sequence with the same validator (Lido) and builder (BeaverBuild) was 11 consecutive slots on March 4th, 2024 (more details with descriptive statistics in the \href{https://colab.research.google.com/drive/1kKM-da6xP7St8puzPuyn1Ndag6a6wsg3#scrollTo=5bje4mIWzELq}{notebook}). Running the same simulation on monthly or total market shares in the time period, the observation shifts to having more longer sequences than expected, however we attribute this to the statistical effect of changing market shares. A detailed analysis can be run in the \href{https://colab.research.google.com/drive/1kKM-da6xP7St8puzPuyn1Ndag6a6wsg3#scrollTo=mz4CTqCQInTv}{notebook} or be provided upon request.

\begin{figure}[H]
    \centering
    \includegraphics[width=0.65\linewidth]{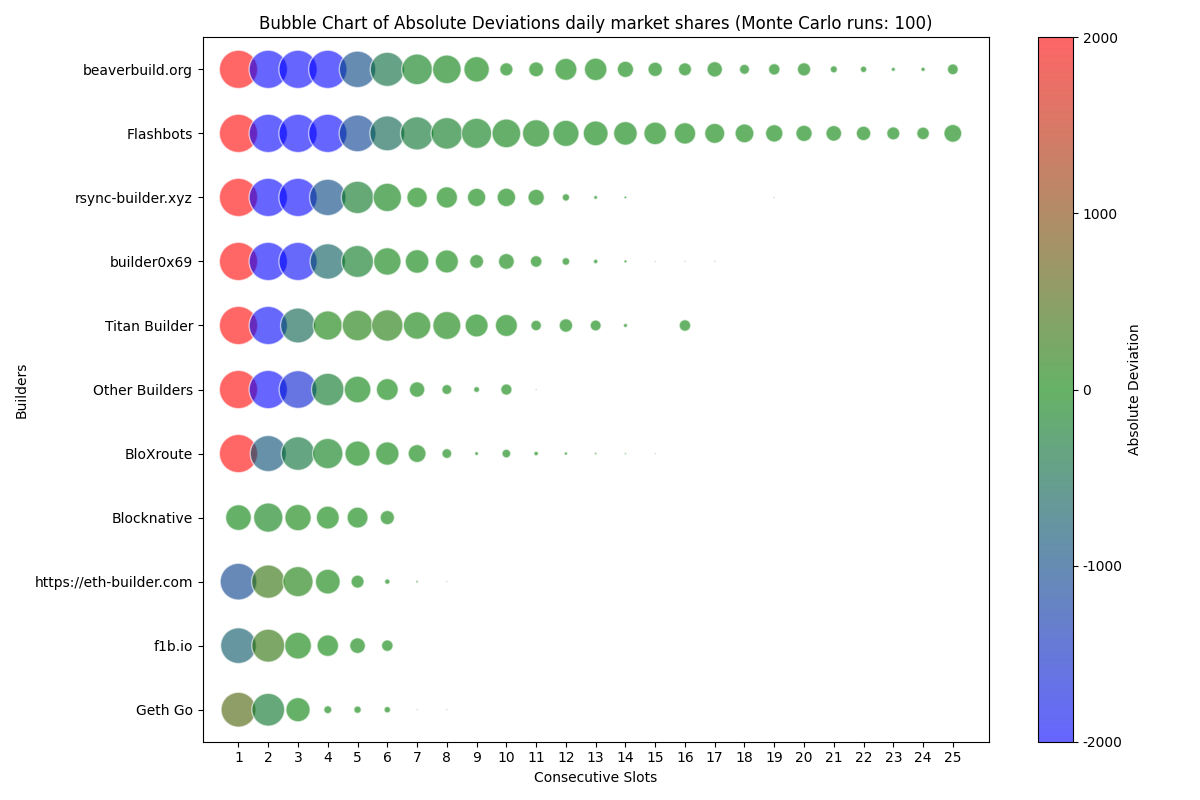}
    \caption{Deviations between expected (Monte Carlo simulation) and observed multi-slot frequencies per builder}
    \label{Fig_2}
\end{figure}

In the next step, to understand this in a more-fine-grained manner, the values are compared for each of the top 10 builders based on market shares. Therefore, for each builder, the difference between expected and observed occurrences of multi-slot sequences are plotted with the size of the bubble indicating the delta in \autoref{Fig_2}. The expected occurrences are based on the results of the Monte Carlo simulation. Red bubbles indicate a positive deviation (more observed slots than expected), while blue indicates a negative deviation. Green dots indicate values in line with the expectation. In \autoref{Fig_2} it is shown in absolute numbers, in the \href{https://colab.research.google.com/drive/1kKM-da6xP7St8puzPuyn1Ndag6a6wsg3#scrollTo=cd07f078-f646-450c-b610-9e91012111f2&line=3&uniqifier=1}{notebook} it can also be seen on a relative scale.

It can be observed in the relative as well as in the absolute deviation that for the top builders there are more single slot sequences than expected with the exception of ETH-Builder, f1b and Blocknative. For multi-slot sequences with two or more slots, almost all top 10 builders have less than expected. This shows that the trend is not limited to singular entities but derives more from the general market structure.

\subsection*{Finding 2: Payments for multi-slot sequences are higher on average than for single slots}
\label{Finding2}

\begin{figure}[H]
    \centering
    \includegraphics[width=0.65\linewidth]{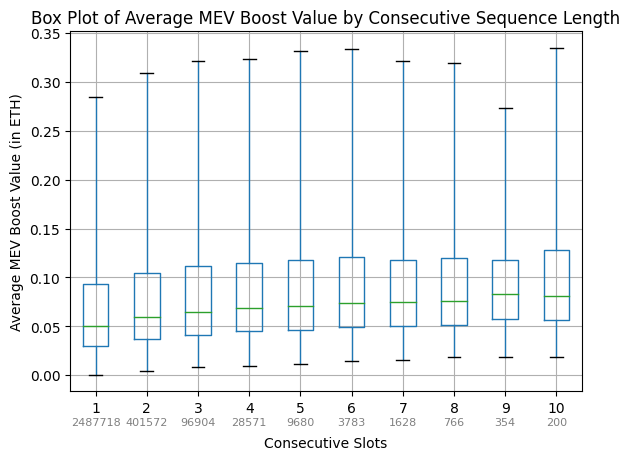}
    \caption{Average MEV-Boost payments per Sequence Length}
    \label{Fig_3}
\end{figure}

To understand if multi-slot sequences are valuable, we looked into MEV-Boost payments and compared single-slot to multi-slot sequences (\autoref{Fig_3}).

\clearpage

It can be observed that in accordance with previous work of \cite{jensen2023multi}, we observe higher average MEV payouts for longer consecutive sequences (from about 0.05 ETH for single slot sequences to around 0.08 ETH for sequences with nine consecutive slots). Note that the gray numbers in \autoref{Fig_3} provide the sample size for each slot length. So it can be observed that the longer the sequence, almost linearly the average MEV-boost payment per slot in the sequence rises. At this stage of the research we can only speculate why this is the case. It could be driven by a higher value in longer consecutive sequences, but also by alternative effects. For example, Julian rightfully pointed out it could also be driven by an increasing intrinsic value for the second highest-bidder due to accumulating MEV in private order flow and the intrinsic valuation of the winning bidder remains constant. Or as Danning suggested, it might be driven by certain types of proprietary order flow (e.g. CEX-DEX arbitrage) being more valuable in certain time periods (e.g. volatile periods) leading to more consecutive sequences as well as higher MEV-Boost payments on average. For a more comprehensive answer and a more in-depth understanding, an analysis on the true block value (builder profits plus proposer payments) and potentially on individual tx level is necessary. We leave this open for future research.

This trend also holds when plotting the average payments for each individual builder. The results on this are shown in the \href{https://colab.research.google.com/drive/1kKM-da6xP7St8puzPuyn1Ndag6a6wsg3#scrollTo=e673f535-1bad-41aa-b617-fcdeee234f01&line=3&uniqifier=1}{notebook}.

\subsection*{Finding 3: Per Slot Payments also increase with longer sequences}

\begin{figure}[H]
    \centering
    \includegraphics[width=0.6\linewidth]{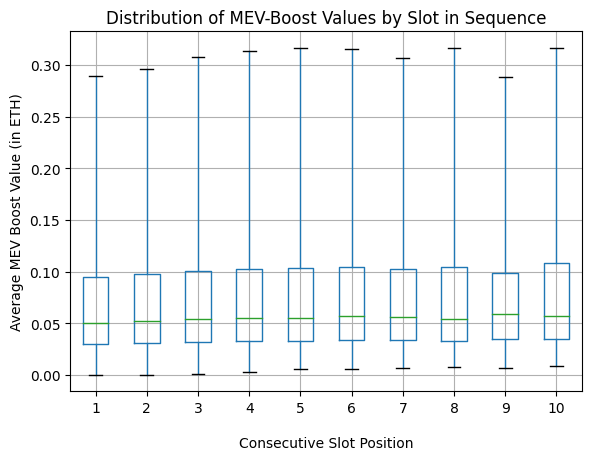}
    \caption{Average MEV-Boost payments per Sequence Position}
    \label{Fig_4}
\end{figure}

Supplementary to the absolute average payment, we also looked into the payment per slot position in longer sequences (\autoref{Fig_4}). E.g. how much was on average paid for the third position in a longer sequence.

Also in the payment per slot analysis a similar trend can be observed, however less prevalent. This suggests that there is slight value in longer sequences, however builders are not willing to bid significantly more for longer consecutive sequences or the first slot after a longer sequence.

This indicates for us that, at least so far, multi-slot strategies are not applied systematically. In this case, we expect builders would need to pay significantly higher values for later slots to ensure to capture the MEV opportunity prepared earlier.

\clearpage

\subsection*{Finding 4: Low auto-correlation between consecutive MEV-Boost payments}

\begin{figure}[H]
    \centering
    \includegraphics[width=0.62\linewidth]{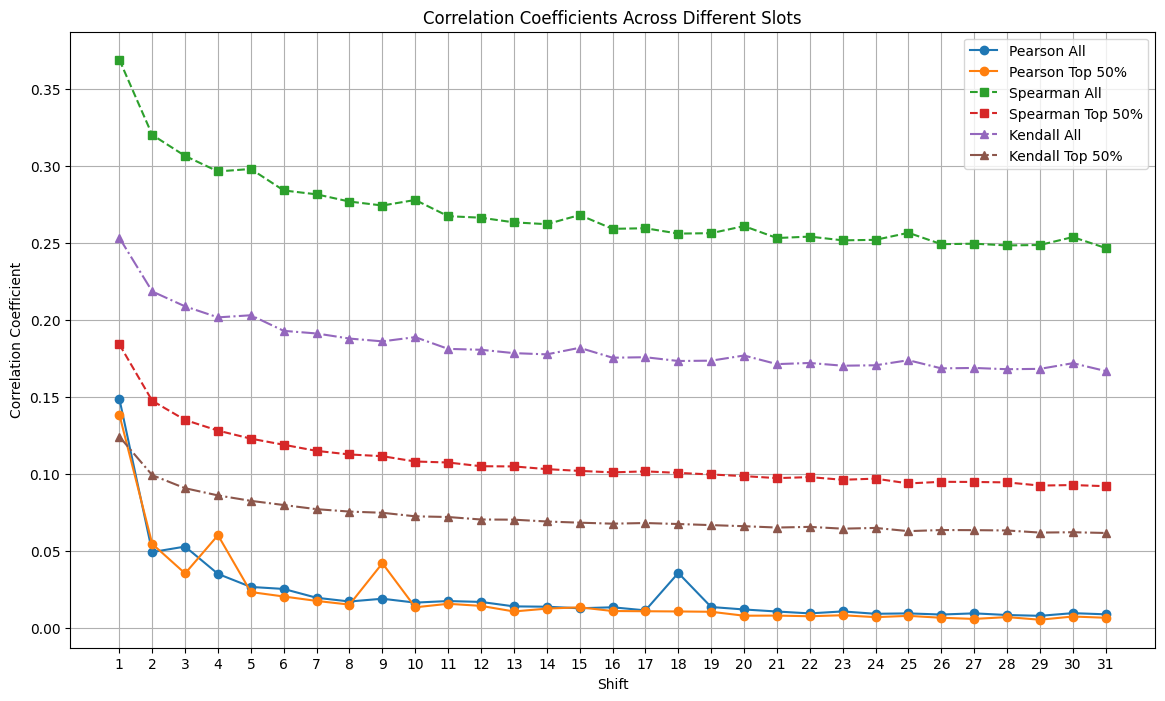}
    \caption{Auto-correlation of MEV-Boost Payments}
    \label{Fig_5}
\end{figure}

We examined auto-correlation in the MEV boost payments to understand if historical MEV data allows us to forecast future MEV and to see if there are low- and high-MEV periods (\autoref{Fig_5}).

Overall, it can be observed that within the first few slots the correlation strongly decreases until an offset of 2 to 3 slots (we tested for Pearson Correlation Coefficient, Spearman’s Rank Correlation Coefficient and Kendall’s Rank Correlation Coefficient). Based on this we can conclude that not more than one to three slots in advance the MEV value can be moderately predicted based on historical data.

Further interesting observations can be made. As expected, the Spearman and Kendall correlation coefficients are significantly higher than the Pearson correlation coefficient, underlining that the data is not following a normal distribution but being skewed and having large outliers. Additionally, it is interesting to note that for the Pearson correlation coefficient, the complete data set and the top 50\% quantile dataset behave similarly, which is not the case for the Spearman and Kendall coefficients. This might be an indicator that the rank ordering for the lower 50\% quantile can be more reliably predicted, further underlying that high MEV values are volatile and spiky, hence difficult to predict.

\subsection*{Finding 5: No indication of builder specialization on low- or high base fee volatility environment}

Previous research (e.g. \cite{gupta2023centralizing} \& \cite{heimbach2024non}) has found that certain builders specialize in low- or high token price volatility environments, with volatility being measured on CEX-price changes. Further, \cite{oz2024wins} observe that different builders have different strategies with some focusing on high-value blocks while others on gaining market share in low-MEV blocks.

Complementary, to determine whether low or high base fee volatility impacts (multi-block) MEV, we analyzed changes in base fee data to identify periods of high volatility. The base fee fluctuations are driven by whether the gas usage in the previous block was below or above the gas target, as defined by \href{https://eips.ethereum.org/EIPS/eip-1559}{EIP-1559}. To identify high volatility environments, we employed two methods: (i) a more naive approach that calculated price changes per slot, classifying the highest and lowest (negative) 10\% of these changes as high volatility periods, with the remaining 80\% of slots being categorized as low volatility. Consequently, high volatility blocks occur following a block with either minimal or significant MEV and/or priority tips. (ii) Secondly, the Garman-Klass volatility \cite{meilijson2008garman} was calculated on an epoch basis, with slots in the top 20\% of GK values designated as high volatility. This approach allows us to examine longer periods characterized by minimal or significant MEV and/or priority tips.

Initial correlation analysis shows only a low correlation between low and high volatile periods and the respective builders (\href{https://en.wikipedia.org/wiki/Cram%C3%A9r%27s_V}{Cramér’s V} for the naive approach 0.0664 and for the Garman-Klass 0.0772). This indicates that there seems to be no builder specialization based on the volatility environment of the base fee. So, it can be observed that in contrast to token price volatility for base price volatility there seems to not have a specialization of builders developed (yet). Further research is needed to elaborate on this first finding.

%*-*-*-*-*-*-*-* [LIMITATIONS] *-*-*-*-*-*-*-*
\section{Limitations}
\label{sec:Limitations}

The research presented here is intended as an initial exploratory analysis of the data rather than a comprehensive study. It is important to note several limitations that affect the scope and conclusions of this analysis. Firstly, it is limited by the considered data set being publicly available MEV-Boost payments data. This leaves out roughly 10\% of non-MEV-Boost facilitated blocks and it does not reflect potential private off-chain agreements. Additionally, the data was partially incomplete and in other parts contained duplicate information (see the \href{https://colab.research.google.com/drive/1kKM-da6xP7St8puzPuyn1Ndag6a6wsg3#scrollTo=0d986969-2492-49ac-ad92-8ff78e2a7fe1&line=2&uniqifier=1}{notebook} for details). Further, missed slots have been excluded so far, a more detailed analysis in the future might focus on the particular effects missed slots have on the subsequent MEV. Lastly, as outlined in the \hyperref[sec:methodology]{methodology} section, using MEV-Boost payments is only a proxy for captured MEV and the competitive metric used in \cite{yang2024decentralization} is only partially applicable for our use case.

As outlined in section \hyperref[Finding2]{Finding 2} it currently can only be speculated about the causation of the increasing average MEV-Boost payouts. Furthermore, running the analysis on the true block value (proposer payment plus builder profits) might generate further insights and solidify the research findings.

On the frequency analysis, the approach contains somewhat a chicken and egg-problem. The Monte Carlo simulation is run on market shares, while the market shares potentially derive from multi-slot sequences. We see a daily time window as an appropriate balance between precision and the need to filter out isolated effects, although this can be critically challenged.

%*-*-*-*-*-*-*-* [CONCLUSIONS] *-*-*-*-*-*-*-*
\section{Conclusions}
\label{sec:Conclusions}

Analyzing block meta-data since the merge, we observe that multi-slot sequences occur less frequently than statistically expected. Further, we observe that the average payments for longer multi-slot sequences increase with the sequence length. Similarly, the payments per slot position in longer sequences also slightly rise. This might indicate that there is generally value in longer consecutive sequences. However, considering the only slight increase in value and the fewer observed multi-slot sequences than expected we so far see no indication of deliberate multi-slot MEV strategies being deployed. Also on individual builder level we currently don’t observe strong deviations from expected distributions. This may also stem from the fact that in the current PBS mechanism, with MEV-Boost operating as a just-in-time (JIT) block auction, creating multi-block MEV opportunities carries inherent risk. This risk arises as creating these opportunities typically requires an upfront investment, and the opportunity might be captured by a competing builder in the next slot, assuming no off-chain collusion between the proposer and builder. This element of risk is a critical factor that could be eliminated by some of the proposed changes to the mechanism (e.g. some APS designs), making it an essential consideration when defining future mechanisms.

%*-*-*-*-*-*-*-* [REFERENCES] *-*-*-*-*-*-*-*
\bibliographystyle{IEEEtran}
\bibliography{References/references}

\end{document}